\documentclass[sigconf, **anonymous**]{acmart}
\AtBeginDocument{%
  }

\setcopyright{acmlicensed}
\copyrightyear{2026}
\acmYear{2026}
\acmDOI{XXXXXXX.XXXXXXX}
\acmConference[WWW '26]{Proceedings of the ACM Web Conference 2024}{April 13--17, 2024}{Dubai, UAE}
\acmBooktitle{Proceedings of the ACM Web Conference 2026 (WWW '26), April 13--17, 2026, Dubai, UAE}
\usepackage{multirow}

\begin{document}

\title{SpatCode: Rotary-based Unified Encoding Framework for Efficient Spatiotemporal Vector Retrieval}


\author{Bingde Hu}
\authornote{Corresponding author.}
\affiliation{%
  \institution{Zhejiang University}
  \city{Hangzhou}
  \country{China}}

\author{Enhao Pan}
\affiliation{%
  \institution{Zhejiang University}
  \city{Hangzhou}
  \country{China}}

\author{Wanjing Zhou}
\affiliation{%
  \institution{Nanjing University}
  \city{Nanjing}
  \country{China}}

\author{Yang Gao}
\authornotemark[1]
\affiliation{%
  \institution{Zhejiang University}
  \city{Hangzhou}
  \country{China}}

\author{Zunlei Feng}
\affiliation{%
  \institution{Zhejiang University}
  \city{Hangzhou}
  \country{China}}

\author{Hao Zhong}
\affiliation{%
  \institution{Zhejiang University}
  \city{Hangzhou}
  \country{China}}

\renewcommand{\shortauthors}{Trovato et al.}

\begin{abstract}
Spatiotemporal vector retrieval has emerged as a critical paradigm in modern information retrieval, enabling efficient access to massive, heterogeneous data that evolve over both time and space. However, existing spatiotemporal retrieval methods are often extensions of conventional vector search systems that rely on external filters or specialized indices to incorporate temporal and spatial constraints, leading to inefficiency, architectural complexity, and limited flexibility in handling heterogeneous modalities.
To overcome these challenges, we present a unified spatiotemporal vector retrieval framework that integrates temporal, spatial, and semantic cues within a coherent similarity space while maintaining scalability and adaptability to continuous data streams. Specifically, we propose (1) a Rotary-based Unified Encoding Method that embeds time and location into rotational position vectors for consistent spatiotemporal representation; (2) a Circular Incremental Update Mechanism that supports efficient sliding-window updates without global re-encoding or index reconstruction; and (3) a Weighted Interest-based Retrieval Algorithm that adaptively balances modality weights for context-aware and personalized retrieval.
Extensive experiments across multiple real-world datasets demonstrate that our framework substantially outperforms state-of-the-art baselines in both retrieval accuracy and efficiency, while maintaining robustness under dynamic data evolution. These results highlight the effectiveness and practicality of the proposed approach for scalable spatiotemporal information retrieval in intelligent systems.
% The exponential growth of multimodal data—ranging from images and text to video and audio—combined with advances in large language models (LLMs) has heightened the demand for intelligent multimodal retrieval systems. However, existing approaches often fall short when handling complex spatiotemporal dynamics across heterogeneous modalities. Some methods focus solely on cross-modal semantic alignment, neglecting spatial and temporal context, while others incorporate such context at the cost of efficiency, scalability, or usability. To bridge this gap, we propose a unified spatiotemporal multimodal retrieval framework that integrates semantic, spatial, and temporal cues within a coherent similarity space, supports high-velocity data updates, and adapts to user-specific retrieval intents. Our framework introduces three key innovations: (1) a Rotary-based Unified Encoding that embeds timestamps and geolocations into rotational position vectors for spatiotemporally-aware representation across modalities; (2) a Circular Incremental Update Mechanism using a sliding-window-based circular buffer to maintain up-to-date indices without expensive global re-encoding; and (3) a Weighted Interest-based Retrieval Algorithm that dynamically adjusts modality weights per query to support personalized, context-aware search. Experiments across representative real-world scenarios demonstrate that our framework achieves superior retrieval accuracy, scalability, and adaptability compared to existing solutions.
\end{abstract}



% \received{}
% \received[revised]{}
% \received[accepted]{}

\maketitle

\section{Introduction}
% Spatiotemporal vector retrieval has emerged as a cutting-edge paradigm in information retrieval, redefining how systems manage and exploit massive, heterogeneous datasets that evolve over both time and space.
Spatiotemporal vector retrieval has emerged as a cutting-edge paradigm in information retrieval, redefining how systems manage and exploit massive, heterogeneous datasets that evolve over both time and space. In particular, with the rapid advancement of large language models, the demand for efficient spatiotemporal retrieval has become increasingly critical for enabling robust context engineering and scalable memory management in intelligent agent systems, such as efficiently and rapidly recalling user interactions from specific time periods or accelerated retrieval of events tied to particular spatial locations. Moreover, in more domain-specific scenarios, spatiotemporal vector retrieval also plays an important role: for example, in emergency management systems, it enables the rapid search of relevant meteorological data based on current time and geological disaster information to support intelligent decision-making; in security systems, it facilitates the fast retrieval of trajectory photos of violation subjects within proximate spatiotemporal ranges based on a single violation snapshot.

\begin{figure}[t]
\centering
\includegraphics[width=0.99\columnwidth]{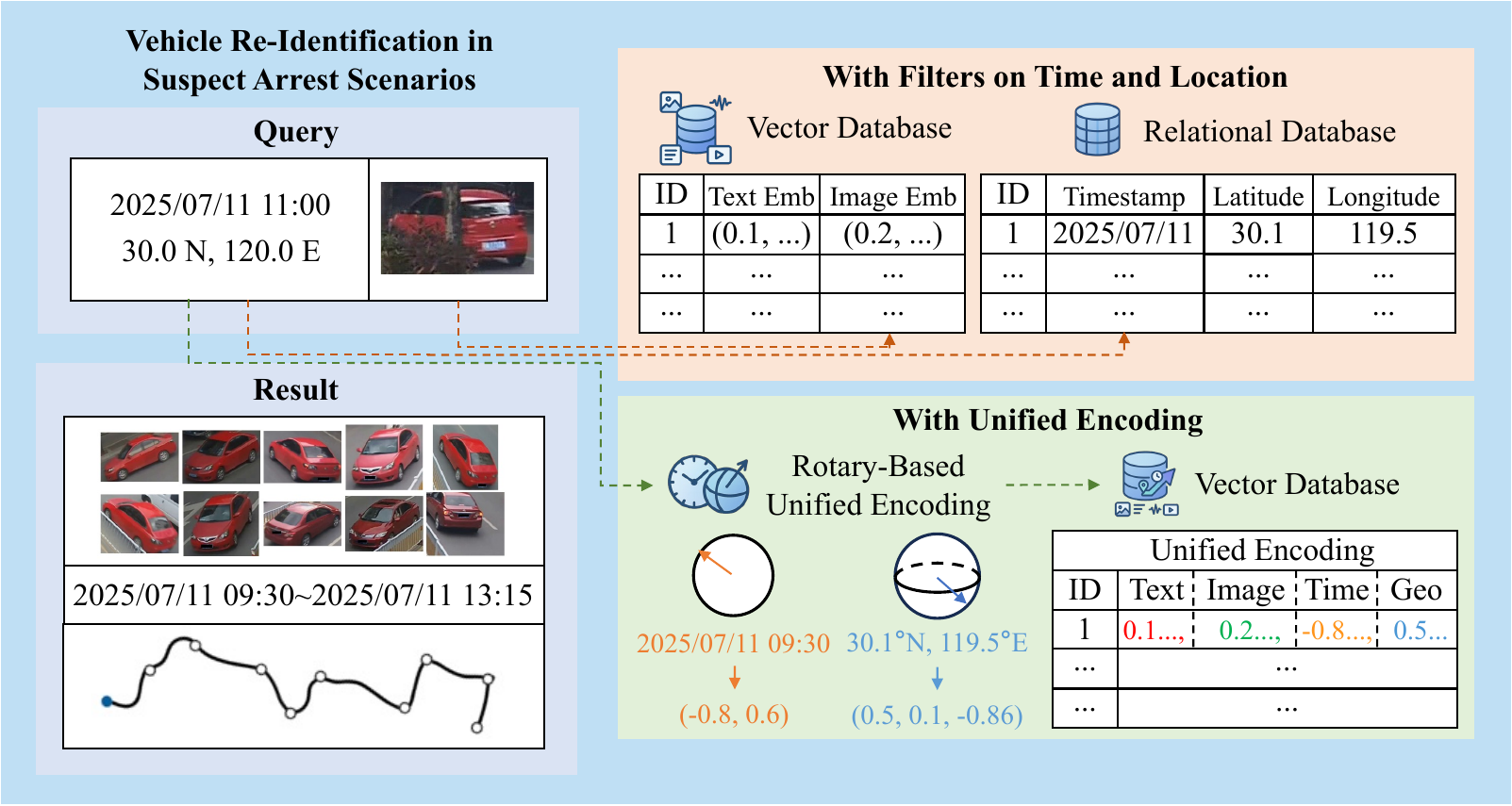}
\caption{Advantages of SpatCode Framework. Traditional spatiotemporal vector retrieval methods rely on additional filters for time and location, increasing system complexity and reducing efficiency. In contrast, unified encoding (SpatCode) embeds spatiotemporal cues into a unified representation, enabling efficient retrieval via a single inner product.}
\label{fig:query_example}
\end{figure}

Despite its importance, existing spatiotemporal retrieval approaches remain largely extensions of conventional vector retrieval methods. As illustrated in Figure \ref{fig:query_example}, to incorporate temporal and spatial attributes, these methods often rely on auxiliary filtering or indexing mechanisms that compromise efficiency, increase system complexity, and limit methodological flexibility.
For example, Milvus \cite{wang2021milvus} and ThalDB \cite{jo2024thalamusdb} introduce temporal or spatial range filters into Approximate Nearest Neighbor (ANN) search, but such hybrid designs introduce additional relational-like structures and extra checks, leading to higher latency and complexity. Similarly, DIGRA \cite{jiang2025digra} enhances vector indexing and dynamic updates through multi-way trees with local proximity graphs, yet it cannot effectively accommodate heterogeneous vectors—multimodal representations that combine features from different sources such as text, image and is restricted to indexing a single attribute (e.g., time or space). More broadly, existing approaches can be categorized into two main families: filter-based extensions of ANN search, which integrate spatiotemporal constraints through post-processing filters but suffer from high query overhead; and graph/tree-based spatiotemporal indices, which support dynamic updates but struggle with heterogeneous vectors and unified modeling. As a result, current solutions either fall short of offering a coherent spatiotemporal modeling framework or impose significant trade-offs in efficiency, scalability, and usability, ultimately limiting their effectiveness in real-world settings.

These limitations raise a fundamental research challenge: \emph{
How can we design a unified spatiotemporal vector retrieval framework that coherently represents heterogeneous modalities with temporal and spatial signals, while ensuring efficiency, scalability, and adaptability to dynamic user needs?}

\begin{figure*}[t]
\centering
\includegraphics[width=0.85\textwidth]{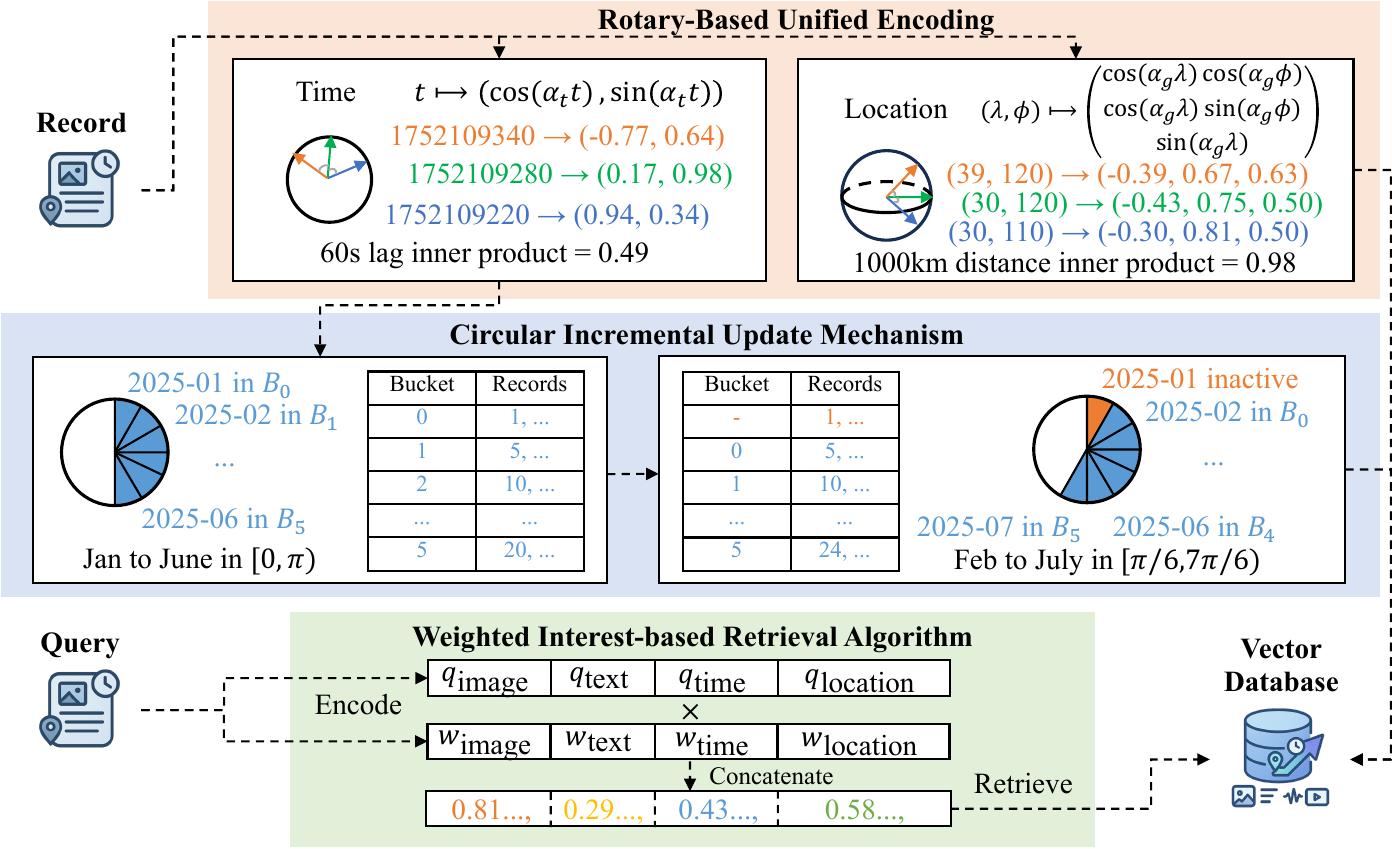}
\caption{SpatCode framework. SpatCode comprises a Rotary-based Unified Encoding Method for integrating spatiotemporal and heterogeneous vectors, a Circular Incremental Update Mechanism for efficient streaming updates, and a Weighted Interest-based Retrieval Algorithm for adaptive, personalized search.}
\label{fig:method}
\end{figure*}

To address this challenge, we propose a unified spatiotemporal vector retrieval framework, illustrated in Figure \ref{fig:method}. Our framework is explicitly designed to integrate spatiotemporal cues into the vector space, support high-velocity data streams, and enable adaptive, user-centric retrieval. 
Specifically, we decompose the problem into three key challenges—representation, evolution, and adaptation—and propose corresponding solutions for each.
First, to coherently represent heterogeneous modalities together with temporal and spatial information, we present a Rotary-based Unified Encoding Method, which embeds time and location into rotational position vectors, ensuring that similarity computations naturally capture relative spatiotemporal relationships.
Second, as data continuously evolve and outdated records accumulate, maintaining efficient and consistent indexing becomes critical. We therefore design a Circular Incremental Update Mechanism, which supports fast sliding-window updates without costly global re-encodings or index reconstructions.
Finally, recognizing that different retrieval tasks emphasize different modalities and temporal contexts, we propose a Weighted Interest-based Retrieval Algorithm that adaptively adjusts modality weights for personalized and context-aware ranking.

In summary, we make the following contributions:
\begin{itemize}
\item Rotary-based Unified Encoding. We propose a novel encoding approach that embeds temporal and spatial signals into rotational position vectors, achieving a unified representation of heterogeneous modalities within a single similarity space. This design enables similarity computations to inherently reflect relative spatiotemporal relationships without the need for external filters or multi-stage processing.
\item Circular Incremental Update Mechanism. To efficiently handle high-velocity and continuously evolving data streams, we develop a sliding-window circular buffer-based update mechanism that incrementally maintains vector indices. This approach ensures low-latency updates and high retrieval consistency without requiring costly global re-encodings or index reconstructions.
\item Weighted Interest-based Retrieval Algorithm. We introduce a flexible retrieval algorithm that supports dynamic, per-query weighting across visual, textual, and spatiotemporal modalities. By adaptively balancing modality relevance, this algorithm enables personalized and context-aware ranking, enhancing retrieval precision across diverse application scenarios.
\item Extensive Experiments and Analyses. We conduct comprehensive experiments on multiple real-world spatiotemporal datasets, demonstrating that our framework consistently outperforms state-of-the-art baselines in both retrieval accuracy and efficiency. Detailed ablation and scalability analyses further validate the effectiveness of each proposed component and the robustness of the method under continuous data evolution.
\end{itemize}

\section{Related Work}
\subsection{Cross-Modal Embeddings}
Early efforts in image-text retrieval aimed to learn a joint embedding space for vision and language \cite{cao2022image}. DeViSE \cite{frome2013devise} mapped images to semantic word vectors via joint training on labels and text for cross-modal search. Later methods enhanced modality alignment with neural encoders and ranking losses. This led to large-scale contrastive models like CLIP \cite{radford2021learning} and ALIGN \cite{jia2021scaling}, which use dual encoders to embed images and text into a shared space. CLIP trained on 400M image-text pairs, optimizing cosine similarity to distinguish matching pairs. ALIGN scaled to 1B noisy pairs, yet its simple contrastive setup still achieved strong retrieval performance.
However, these models primarily focus on static image-text alignment and do not account for spatiotemporal or contextual dynamics in multimodal retrieval. In contrast, our approach extends this paradigm by incorporating temporal and geographic embeddings into a unified representation, enabling adaptive retrieval across heterogeneous modalities.

\subsection{Spatiotemporal Encoding and Indexing}
Many retrieval tasks require filtering results by time and location in addition to semantic relevance \cite{li2023progressive}, yet integrating spatiotemporal constraints into vector-based search remains challenging. Traditional indexes like R-trees (spatial) and B-trees (temporal) \cite{yufei2000mv3r} handle structured attributes but do not support high-dimensional similarity on unstructured content. Recent efforts explore time-aware vector search \cite{zeinalipour2006distributed, tedjopurnomo2021similar, fang2022spatio}, and one common approach is to build multiple specialized indexes. For instance, SIEVE \cite{li2025sieve} uses sub-indexes tailored to different filters, routing queries accordingly. Geospatially, it's common to pre-filter by range or geohash using latitude and longitude before applying vector similarity \cite{patroumpas2020similarity}. However, no unified method yet elegantly combines spatiotemporal constraints with semantic similarity in a single embedding or index. Our method addresses this gap by introducing a unified spatiotemporal embedding framework that integrates time, location, and semantic features within a single vector space. Unlike prior multi-index or pre-filtering approaches, it enables direct similarity computation under joint spatiotemporal-semantic contexts, achieving both representational coherence and retrieval efficiency.

\subsection{ANN Algorithms}
The backbone of modern vector retrieval is Approximate Nearest Neighbor (ANN) search, which mainly includes quantization-, graph-, and tree-based methods \cite{li2022deep}. Quantization techniques like Product Quantization (PQ) compress vectors by partitioning them into subspaces and quantizing each one \cite{jegou2010product}, enabling efficient distance estimation with reduced memory. Graph-based methods such as HNSW \cite{malkov2018efficient} build multi-layer navigable small-world graphs and achieve strong recall-speed trade-offs via hierarchical greedy search. Tree-based and hardware-optimized approaches also play roles in specific scenarios \cite{aumuller2023recent}. Each has strengths: quantization suits memory-constrained setups, graphs offer high recall, and trees perform well in low dimensions or as filters. Our method builds upon these ANN foundations but extends them to handle spatiotemporal and heterogeneous vector representations.

% \subsection{Vector Databases and Filtering}
% Vector database systems like Milvus \cite{wang2021milvus,han2023comprehensive} build on ANN algorithms to manage large-scale embeddings with dynamic updates, distributed storage, and hybrid queries over vectors and attributes. Milvus supports modular ANN indexes and metadata filtering (e.g., tags, timestamps), though naive pre-/post-filtering can degrade performance. To address this, systems like SIEVE \cite{li2025sieve} partition data by filters to improve speed, at the cost of larger indexes and maintenance complexity.
% Despite such advances, hybrid search remains challenging. Researchers apply segmented indexing or re-ranking without unified query support \cite{pan2024survey}. Moreover, most vector DBs process multimodal data separately, lacking integrated representations that embed both modalities and metadata.

\section{Preliminary}

In this section, we present the essential terminology and formally define the spatiotemporal retrieval problem.

\paragraph{\textbf{Record Set}}
Let the database be $\mathcal{D}=\{x_n\}_{n=1}^N$. Each record is written as
$$
x_n=\{f_n,t_n,(\lambda_n,\phi_n)\},
$$
where $f_n\in\mathbb{R}^d$ denotes the content features of the $n$-th item (potentially comprising $m$ modality-specific embeddings $f_n^{(i)}\in\mathbb{R}^{d_i}$), $t_n\in\mathbb{R}$ is its acquisition time on a monotonically increasing global clock and $(\lambda_n,\phi_n)$ are its geodetic latitude and longitude coordinates.

\paragraph{\textbf{Query}}
At query time, the user provides modality-specific cues together with a set of user-specified weights. We write
$$
q=\left(\{q^{(i)}\in\mathbb{R}^{d_i}\}_{i\in\mathcal{M}_{\text{content}}},q_{\text{time}}=t_q,q_{\text{location}}=(\lambda_q,\phi_q)\right),
$$
together with $w=(w_i)_{i\in\mathcal{M}}$. Here, $\mathcal{M}_{\text{content}}\subseteq\mathcal{M}$ indexes all non-spatiotemporal content modalities (e.g., image, text, audio) in a generic way, while two distinguished modalities capture time $t_q$ and geolocation $(\lambda_q,\phi_q)$.

\paragraph{\textbf{Problem Statement}}
Given $\mathcal{D}$, a query $q$, and an integer $k\ge1$, the goal is to return a set of items of size $k$
$$
\mathcal{N}_k(q)\subset\mathcal{D},\ |\mathcal{N}_k(q)|=k,
$$
that maximizes a similarity score $S(q,x)$ capturing semantic or content alignment, temporal proximity to $q_{\text{time}}$, and spatial proximity to $q_{\text{location}}$. Formally, the top-$k$ set is
$$
\mathcal{N}_k(q)=\arg\text{top-}k_{x\in\mathcal{D}}\ S(q,x).
$$

\paragraph{\textbf{Performance Metric}}
Let $R_k$ be the retrieved set (top-$k$ returned by the system) and $G_k$ the ground-truth top-$k$ under $S(q,x)$ for the given query-specific weights. We report:
$$
\text{recall}@k=\frac{|R_k\cap G_k|}{k}.
$$

\section{Method}
Figure \ref{fig:method} gives an overview of the three components of SpatCode and their interactions in the retrieval pipeline. In the following three sections, we detail its core components: the Rotary-based Unified Encoding Method, the Circular Incremental Update Mechanism, and the Weighted Interest-based Retrieval Algorithm. For the precision and complexity analysis of the method, see Appendix.

\subsection{Rotary-based Unified Encoding}
% Spatiotemporal data often contain three complementary signals—content, time, and location—that must be jointly understood for effective retrieval. However, conventional vector representations typically treat temporal and spatial metadata as external filters or separate dimensions, which leads to inefficient and inconsistent similarity computation. To overcome this, we introduce a **Rotary-based Unified Encoding** that embeds all three aspects—semantic, temporal, and geographic—into a single, coherent vector space. This design enables spatiotemporal similarity to be computed directly through standard dot-products, while preserving the relative relationships between events in both time and space. Conceptually, this encoder translates time and location into smooth rotations on a hypersphere, so that temporal lag and spatial distance correspond naturally to angular differences in the embedding space.

A central challenge in spatiotemporal retrieval is to establish a consistent representation that seamlessly fuses heterogeneous feature vectors with their temporal and spatial attributes, while preserving the semantic relationships between data items. Instead of relying on spatiotemporal post-processing filters or graph/tree-based spatiotemporal indices, we propose a \emph{Rotary-based Unified Encoding} method, which encodes time and geographic information directly into the vector space using periodic angular functions. This design allows the model to capture relative temporal and spatial relationships through simple cosine similarity, enabling compatibility with standard vector retrieval methods such as HNSW without any specialized spatiotemporal index.

We represent a data item as $x_n=\{f_n,t_n,(\lambda_n,\phi_n)\}$, where $f_n$ is the feature vector, $t_n\in\mathbb{R}$ is the timestamp measured in seconds and $(\lambda_n,\phi_n)$ is the geographic coordinate, where $\lambda_n\in[-\pi/2,\pi/2]$ is the latitude and $\phi_n\in[-\pi,\pi)$ is the longitude expressed in radians. The central design objective is to transform these three scalars $(t_n, \lambda_n,\phi_n)$ into unit norm vectors such that: \emph{(i) the inner product between any two time encodings depends only on their temporal lag; (ii) the inner product between any two location encodings depends only on their great-circle separation; (iii) the time encoding is governed by a tunable scale parameter that balances representation range against resolution.}
Under these constraints, similarity computation within standard vector databases simplifies to cosine similarity in $\mathbb{R}^d$, while still faithfully capturing relative chronology and spatial proximity.

The unified encoder is decomposed into its temporal and spatial components. For the temporal dimension $t_n$, it should be embedded as a unit vector $(x, y)\in \mathbb{R}^2$ on the unit circle (satisfies $x^2+y^2=1$).
Thus, fix a temporal scale parameter $\alpha_t>0$ that converts a physical time interval (for example, one second or one day) to an angular displacement. Define the mapping
$$
E_t: t_n\mapsto(\cos(\alpha_tt_n),\sin(\alpha_tt_n)).
$$
Let $t_i,t_j$ be two timestamps and set $\Delta t=t_i-t_j$. The cosine similarity of their embeddings is simply
$$
\langle E_t(t_i),E_t(t_j)\rangle=\cos(\alpha_t\Delta t),
$$
which depends exclusively on the lag $\Delta t$ and is indifferent to the absolute epoch. Consequently, any pair of temporal events separated by the same duration exhibits identical similarity, an invariance that aligns with retrieval semantics in which recency difference rather than absolute time governs relevance. The parameter $\alpha_t$ dictates the periodicity of the mapping: a full rotation of $2\pi$ is completed after an interval of $2\pi/\alpha_t$. To ensure monotonicity, one chooses $\alpha_t$ so that the maximum interval of interest $\Delta t_{\max}$ satisfies $\alpha_t\Delta t_{\max}\le\pi$. Under this regime, distinct lags up to $\Delta t_{\max}$ are uniquely represented, and the similarity $\cos(\alpha_t\Delta t)$ is monotonically decreasing in $|\Delta t|$ on $[0,\Delta t_{\max}]$.

The complementary mapping for geography is described below, which preserves relative separation on the globe. Spatial encoding requires a mapping that captures two-dimensional surface geometry. We treat the Earth as a unit radius sphere and interpret $(\lambda_n,\phi_n)$ as spherical coordinates. The mapping
$$
E_g:(\lambda_n,\phi_n)\mapsto
\begin{pmatrix}
\cos(\lambda_n)\cos(\phi_n)\\
\cos(\lambda_n)\sin(\phi_n)\\
\sin(\lambda_n)
\end{pmatrix}
$$
projects coordinates to the unit sphere $(x,y,z)\in\mathbb{R}^3$ (satisfies $x^2+y^2+z^2=1$). Let $(\lambda_i,\phi_i)$ and $(\lambda_j,\phi_j)$ be two locations. Their embedding inner product satisfies
\begin{align*}
&\langle E_g(\lambda_i,\phi_i),E_g(\lambda_j,\phi_j)\rangle=\cos(\delta),\\
&\delta=\arccos(\sin(\lambda_i)\sin(\lambda_j)\\
&+\cos(\lambda_i)\cos(\lambda_j)\cos(\phi_i-\phi_j)),
\end{align*}
where $\delta$ is the great-circle distance of the points on the sphere. Importantly, if two pairs of points share the same great-circle separation, then $\delta$ is identical for both, and hence their inner products coincide. A detailed discussion of the precision of the spatiotemporal encoding is provided in Appendix A.

The elegance of the two encodings lies in their common mathematical form. Each converts a scalar or pair of scalars into a unit norm vector via the exponential map $x\mapsto(\cos x,\sin x)$ (for one dimension) or its spherical analogue (for two dimensions). The inner product is therefore an immediate proxy for offset. This property, together with norm preservation, ensures compatibility with ANN indices such as HNSW or product quantization, whose heuristics typically assume fixed-length vectors and dot-product or cosine scoring.

\subsection{Circular Incremental Update Mechanism}
With the temporal encoder established, we next address the dynamic maintenance of the spatiotemporal index under continuous data arrivals. In real-world streaming environments, where records evolve rapidly over time, outdated embeddings can quickly accumulate and degrade retrieval accuracy. To ensure the index remains both fresh and computationally stable, we introduce a Circular Incremental Update Mechanism, which maintains temporal consistency through a sliding-window-based circular buffer. This design enables incremental insertion and expiration of records without global re-encoding or structural reconstruction, thereby preserving retrieval quality while sustaining high update throughput.

This mechanism is motivated by the temporal periodicity of the encoding, which induces a finite horizon that must be managed explicitly. Because the cosine similarity of two time embeddings depends solely on the phase difference $\alpha_t|t_i-t_j|$, accurate temporal ordering is preserved only for phase separations strictly smaller than $\pi$; beyond that point, the circular map folds back and aliases incomparable epochs. Consequently, any deployment must have a finite horizon length $\Delta t_h\le\pi/\alpha_t$ within which the pairwise ordering is unambiguous. A naive solution would periodically re-encode all stored vectors with a fresh origin as the wall clock advances; such wholesale recomputation, however, scales linearly in the database size and incurs intolerable interruption for high-velocity streams. We therefore design a circular incremental update mechanism that maintains an exact sliding window of length $\Delta t_h$ inside a fixed ANN index while avoiding any in-place modification of previously inserted vectors.

Specifically, let $x_n=\{f_n,t_n,(\lambda_n,\phi_n)\}$ denote a data item in a continuously arriving sequence of spatiotemporal records, where $f_n$ is the feature vector and $t_n\in\mathbb{R}$ is the acquisition time measured in a monotonically increasing global clock. The temporal encoder $E_t$ introduced earlier maps any timestamp $t_n$ to the unit circle by $E_t(t_n)=(\cos(\alpha_tt_n),\sin(\alpha_tt_n))$. Let the window length be a rational multiple of a base unit $\tau$ (for example, one calendar month) so that $\Delta t_h=L\tau$ for some small integer $L$. Define the phase step $\Delta\theta=\alpha_t\tau$ and, without loss of generality, choose $\alpha_t$ such that $L\Delta\theta=\pi$. The entire non-aliasing interval thus occupies the semicircle $[0,\pi)$ in phase space. The database is partitioned into $L$ disjoint phase buckets $B_0,\dots,B_{L-1}$; bucket $B_k$ stores exactly the items whose timestamp lies in the half-open interval $[T_0+k\tau,T_0+(k+1)\tau)$, where $T_0$ is an arbitrarily chosen epoch boundary. Each bucket is implemented as a contiguous range of internal identifiers within the ANN index and, additionally, as an external manifest that maps bucket identity to the list of record IDs it currently occupies. The manifests are lightweight: a bucket header stores two integers (start and end offsets) and a pointer to a variable-length vector of record IDs.

Insertion proceeds as follows. When a new record $(x,t)$ arrives, calculate its phase value $\theta=\alpha_tt \bmod2\pi$. Because $t$ is non-decreasing, $\theta$ drifts uniformly forward, completing a full revolution every period $2\pi/\alpha_t=2\Delta t_h$. To ensure that at any moment only the most recent $\Delta t_h$ worth of data populate the semicircle $[0,\pi)$, we apply a phase shift $\varphi$ which monotonically increases and is updated once per unit step $\tau$. Specifically, define the active mapping
$$
\theta^*=(\theta-\varphi)\bmod2\pi,
$$
and assign the record to bucket $B_k$ with $k=\lfloor\theta^*/\Delta\theta\rfloor$ if $\theta^*\in[0,\pi)$; otherwise the record is outside the sliding window and should be discarded. The shift $\varphi$ takes discrete values in $\{0,\Delta\theta,2\Delta\theta,\dots,(2L-1)\Delta\theta\}$. At the beginning of each interval $[T_0+n\tau,T_0+(n+1)\tau)$, we set $\varphi\leftarrow n\Delta\theta$. Geometrically, this rotates the semicircular aperture of length $\pi$ synchronously with absolute time, so that exactly the newest $L$ unit intervals fall inside it. Crucially, shifting $\varphi$ does not alter any previously stored phase coordinates; the reinterpretation is purely logical and affects only bucket addressing.

Deletion leverages the same periodicity. When $\varphi$ increases from $(n-1)\Delta\theta$ to $n\Delta\theta$, bucket $B_{(n-1)\bmod L}$ becomes obsolete because its contents correspond to times more than $\Delta t_h$ old. We reclaim that bucket by (i) marking its entries as inactive in the ANN index so that they are ignored during similarity search and (ii) erasing its manifest. No global rebuild is required; the index topology remains valid because inactive nodes are simply skipped during traversal. A comprehensive analysis of the algorithm's complexity is presented in Appendix B.

The correctness of similarity evaluation under the sliding aperture follows from the trigonometric identity
\begin{align*}
\langle E_t(t_i),E_t(t_j)\rangle&=\cos(\alpha_t(t_i-t_j))\\
&=\cos((\theta_i-\varphi)-(\theta_j-\varphi))\\
&=\cos(\theta_i^*-\theta_j^*),
\end{align*}
which is invariant under simultaneous rotation of both phases by $\varphi$. Consequently, the inner product between any two live records (that is, those whose phase after shifting lies in $[0,\pi)$) equals their original similarity, while cross-terms between a live record and an expired record are irrelevant because expired entries are flagged as inactive. The mechanism therefore preserves rank equivalence with respect to the intended temporal metric.

In essence, the circular incremental update mechanism leverages the rotational symmetry of encodings, turning window maintenance into simple pointer rotation and bucket aging. This avoids re-encoding vectors or rewiring graph edges. It combines the efficiency of ring buffers, the locality of batch deletion, and the robustness of cosine-based ANN, offering a scalable solution for real-time spatiotemporal retrieval.

\subsection{Weighted Interest-based Retrieval Algorithm}
As the final component of our framework, we shift the focus from data maintenance to query behavior, introducing a lightweight yet expressive mechanism that adapts retrieval ranking to user intent. The Weighted Interest-based Retrieval Algorithm is designed to dynamically adjust the relative importance of heterogeneous modalities (e.g., text, image, and spatiotemporal features), enabling personalized, context-aware retrieval atop the unified index constructed in previous stages.
Unlike the preceding components, which ensure representational consistency and temporal freshness, this module operates at query time, modulating similarity computation without modifying stored embeddings. Through a simple but principled weighting formulation, it closes the methodological loop of the system—linking encoding, index maintenance, and interactive retrieval adaptation into a coherent end-to-end framework.

% Finally, we turn from data maintenance to query behavior, introducing a simple weighting scheme that adapts ranking to user intent. Weighted Interest-based Retrieval Algorithm is developed to dynamically adjust the weighting of different modalities (e.g., image, text, spatiotemporal), enabling personalized and context-aware multimodal retrieval.

In a heterogeneous vector retrieval, typical modalities include textual semantics, visual semantics, audio cues, and the spatiotemporal vector described in the previous section, and the user may attach different importance to each modality. To meet this need, we design a weighted interest-based retrieval algorithm. Specifically, for a data item $x_n=\{f_n,t_n,(\lambda_n,\phi_n)\}$, $f_n$ is an L2-normalized feature vector, $E_t(t_n)$ maps timestamp $t_n$ to an L2-normalized vector on the unit circle and $E_g(\lambda_n,\phi_n)$ maps latitude $\lambda_n$ and longitude $\phi_n$ to an L2-normalized vector on the unit sphere. We generalize a data item as an ordered collection of $m$ L2-normalized sub-embeddings $(v_j^{(1)},\dots,v_j^{(m)})$, with $v_j^{(i)}\in\mathbb{R}^{d_i}$ and $\|v_j^{(i)}\|_2=1$. The framework imposes no constraints on $m$ or on the individual dimensionalities $d_i$. For efficient storage and retrieval, these sub-embeddings are concatenated offline into a single composite embedding
$$
\tilde{v}_j=[v_j^{(1)};\dots;v_j^{(m)}]\in\mathbb{R}^d,\ d=\sum_{i=1}^md_i,
$$
followed by a global L2 normalization $\tilde{v}_j\leftarrow\tilde{v}_j/\|\tilde{v}_j\|_2$. Because the original components are already unit length, the norm of $\tilde{v}_j$ equals $\sqrt m$, so the post-normalization divides by $\sqrt m$; this constant rescaling does not affect the ordering of the inner product, but prepares the vectors for cosine-based ANN indices whose heuristics assume unit norms.

The query interface scales per-modality cues by an interest profile, producing a single compatible query vector. At query time, the user provides a set of modality-specific cues $(q^{(1)},\dots,q^{(m)})$ together with a non-negative weight vector $w=(w_1,\dots,w_m)\in\mathbb{R}^m$. Each $q^{(i)}\in\mathbb{R}^{d_i}$ is L2-normalized, either by construction when the query modality embedding network outputs unit vectors or by explicit normalization in the retrieval front-end. The interest profile $w$ is interpreted as a linear importance prior: larger $w_i$ magnifies the contribution of modality $i$ to the overall similarity, while $w_i=0$ effectively removes that modality from consideration. To assemble a single query vector compatible with the pre-concatenated database, we perform a simple element-wise scaling followed by concatenation:
$$
\tilde{q}=[w_1q^{(1)};\dots;w_mq^{(m)}]\in\mathbb{R}^d.
$$
Note that each block retains its original dimensionality $d_i$, avoiding any learned projection layers; the method remains parameter-free at inference. The norm of $\tilde{q}$ is $\|\tilde{q}\|_2=\sqrt{\sum_{i=1}^mw_i^2}$, which is constant across database candidates for fixed $w$ and therefore immaterial to ranking when cosine similarity is used. In practice, either no further normalization is applied, which yields a dot product scoring, or $\tilde{q}$ is divided by its norm to restore the unit-length assumption of the index. Both variants yield identical orderings. Multiplying all scores by the same positive constant does not affect ranking.

A scoring rule ties the weighted query to the concatenated database embeddings, completing the retrieval formulation. The relevance score between the query and a database element $x_j$ is the inner product
$$
S(x_j,\tilde{q})=\langle\tilde{v}_j,\tilde{q}\rangle=\sum_{i=1}^mw_i\langle v_j^{(i)},q^{(i)}\rangle,
$$
demonstrating algebraically that weighted concatenation recovers the canonical weighted sum of per-field similarities. Because the database vectors are embedded once and for all, a single ANN search in $\mathbb{R}^d$ suffices to obtain the top-$k$ items under the weighted score, eliminating the need for $m$ independent searches followed by set union or re-ranking.

By concatenating L2-normalized modality embeddings and scaling each query block with user-defined weights, the algorithm converts multi-criteria ranking into a single cosine similarity search that exactly implements the desired weighted sum; this preserves index efficiency through unit norm invariance while allowing intuitive control over modality importance. Because only one graph traversal, heap, and scoring pipeline is invoked, rather than $m$ parallel instances, latency is reduced. Recall improves as moderately similar items across modalities surface by their aggregate score.

\section{Experiment}
\subsection{Experiment Setting}
\subsubsection{Dataset Construction and Query Setup}

We constructed a synthetic yet realistic \textbf{Shopping} dataset by pairing CelebA faces \cite{liu2015deep} ($\sim$200K images, 10177 identities) with SQID product images \cite{ghossein2024shopping} ($\sim$180K items). From these, we sampled $\sim$150K records, each containing a face, product, timestamp (uniformly from 2024-2025), and geolocation (lat 29.18°-30.57° N, lon 118.33°-120.62° E). We also evaluate our method on four representative datasets, covering diverse modalities and varying spatiotemporal characteristics.
The \textbf{Craigslist} dataset \cite{jo2024thalamusdb} ($\sim$15K records) contains multimodal records with images, text, timestamps, and locations. We geocode locations and extract 512-D CLIP embeddings for both image and text.
The \textbf{Netflix} dataset \cite{jo2024thalamusdb} ($\sim$18K records) consists of dual-text records (title and synopsis), with synthetic timestamps and GPS coordinates. Both text fields are encoded using CLIP.
The \textbf{Bridge} dataset \cite{mundt2019meta} ($\sim$4K records) includes image records composed of five common defect categories found in 30 unique bridges. We synthesized timestamps and GPS coordinates.
The \textbf{VeRi} dataset \cite{liu2016large} ($\sim$38K records) contains images of 776 vehicles captured by 20 cameras labeled with sufficient license plates and spatiotemporal information.

These datasets span heterogeneous modality combinations and spatiotemporal patterns, enabling comprehensive evaluation under realistic spatiotemporal retrieval conditions.

\subsubsection{Metrics}
For each method, we measure insert and top-100 query latency to assess trade-offs in speed and accuracy under streaming updates. Relevance is defined by the top-100 items with highest weighted similarity $S(q, x)$ using query-specific weights.

\subsubsection{Baseline Methods and Implementation}
We compare our proposed method against three baselines: \emph{ThalDB} stores image paths with timestamps and coordinates in a relational table, using SQL queries with text predicates and scalar filters for threshold-based retrieval. \emph{Milvus Scalar Filtered Search} encodes each record as a 1024-d CLIP vector with scalar time and location fields, applying strict spatiotemporal filtering before a single ANN search. \emph{Milvus Multi-Vector Hybrid Search} indexes face, product, time, and location embeddings separately, performs four parallel ANN searches, and re-ranks results with an RRF ranker to support soft spatiotemporal constraints at the cost of higher query overhead.

We implement SpatCode based on Milvus and evaluate the above methods on a server equipped with 1 TB of RAM and Intel Xeon Platinum 8358P CPU.

\begin{table*}[t]
\caption{Average latency (ms) for inserting and querying a single record across the evaluated methods.}
\label{tab:latency}
\centering
\small
\begin{tabular}{lcccccccccc}
\toprule
\multirow{2}{*}{Method} & \multicolumn{2}{c}{Shopping} & \multicolumn{2}{c}{Craigslist} & \multicolumn{2}{c}{Netflix} & \multicolumn{2}{c}{Bridge} & \multicolumn{2}{c}{VeRi} \\
\cmidrule(lr){2-3} \cmidrule(lr){4-5} \cmidrule(lr){6-7} \cmidrule(lr){8-9} \cmidrule(lr){10-11}
 & Insert & Query & Insert & Query & Insert & Query & Insert & Query & Insert & Query \\
\midrule
ThalDB & 0.0501 & $>$1000 & 0.520 & $>$1000 & 0.0343 & $>$1000 & 0.0022 & $>$1000 & 0.0415 & $>$1000 \\
Filtered & 3.88 & 143 & 3.78 & 11.5 & 3.95 & 61.7 & 3.80 & 17.9 & 3.77 & 32.54 \\
Hybrid & 3.82 & 8.07 & 3.90 & 5.70 & 3.97 & 5.89 & 3.79 & 6.95 & 3.70 & 5.56 \\
\midrule
\textbf{SpatCode} & \textbf{3.69} & \textbf{5.24} & \textbf{3.62} & \textbf{4.03} & \textbf{3.84} & \textbf{4.21} & \textbf{3.63} & \textbf{3.81} & \textbf{3.60} & \textbf{3.87} \\
\bottomrule
\end{tabular}
\end{table*}

% \begin{table*}[t]
% \caption{Retrieval effectiveness for the evaluated methods at various top-$k$ thresholds with HNSW $e_f=100$.}
% \label{tab:retrieval}
% \centering
% \small
% \begin{tabular}{lcccccccc}
% \toprule
% Method & recall@$k$ & Shopping & Craigslist & Netflix & Bridge & VeRi \\
% \midrule
% \multirow{4}{*}{ThalDB}
%   & 1   & 0.276 & 0.356 & 0.284 & 0.284 & 0.106 \\
%   & 10  & 0.358 & 0.429 & 0.399 & 0.290 & 0.116 \\
%   & 50  & 0.333 & 0.429 & 0.355 & 0.200 & 0.109 \\
%   & 100 & 0.316 & 0.399 & 0.314 & 0.159 & 0.101 \\
% \midrule
% \multirow{4}{*}{Filtered}
%   & 1   & 0.316 & 0.452 & 0.296 & 0.408 & 0.134 \\
%   & 10  & 0.414 & 0.540 & 0.423 & 0.449 & 0.142 \\
%   & 50  & 0.415 & 0.556 & 0.397 & 0.327 & 0.139 \\
%   & 100 & 0.408 & 0.532 & 0.362 & 0.269 & 0.129 \\
% \midrule
% \multirow{4}{*}{Hybrid}
%   & 1   & 0.000 & 0.042 & 0.026 & 0.094 & 0.016 \\
%   & 10  & 0.006 & 0.116 & 0.115 & 0.172 & 0.100 \\
%   & 50  & 0.011 & 0.144 & 0.077 & 0.137 & 0.096 \\
%   & 100 & 0.016 & 0.149 & 0.078 & 0.157 & 0.100 \\
% \midrule
% \multirow{4}{*}{\textbf{SpatCode}}
%   & 1   & \textbf{ } & \textbf{0.876} & \textbf{0.696} & \textbf{0.944} & \textbf{0.934} \\
%   & 10  & \textbf{ } & \textbf{0.908} & \textbf{0.865} & \textbf{0.958} & \textbf{0.961} \\
%   & 50  & \textbf{ } & \textbf{0.900} & \textbf{0.899} & \textbf{0.965} & \textbf{0.950} \\
%   & 100 & \textbf{ } & \textbf{0.879} & \textbf{0.894} & \textbf{0.966} & \textbf{0.941} \\
% \bottomrule
% \end{tabular}
% \end{table*}

\begin{table}[t]
\centering
\caption{Retrieval effectiveness for evaluated methods at various top-$k$ thresholds with HNSW $e_f=100$.}
\label{tab:retrieval}
\resizebox{\columnwidth}{!}{%
\begin{tabular}{lcccccc}
\toprule
Method & recall@$k$ & Shopping & Craigslist & Netflix & Bridge & VeRi \\
\midrule
\multirow{4}{*}{ThalDB}
  & 1   & 0.146 & 0.378 & 0.356 & 0.318 & 0.120 \\
  & 10  & 0.187 & 0.455 & 0.513 & 0.291 & 0.129 \\
  & 50  & 0.184 & 0.500 & 0.554 & 0.204 & 0.131 \\
  & 100 & 0.180 & 0.500 & 0.549 & 0.162 & 0.127 \\
\midrule
\multirow{4}{*}{Filtered}
  & 1   & 0.188 & 0.476 & 0.376 & 0.446 & 0.150 \\
  & 10  & 0.243 & 0.568 & 0.547 & 0.452 & 0.158 \\
  & 50  & 0.257 & 0.625 & 0.601 & 0.335 & 0.165 \\
  & 100 & 0.254 & 0.637 & 0.608 & 0.274 & 0.162 \\
\midrule
\multirow{4}{*}{Hybrid}
  & 1   & 0.010 & 0.052 & 0.036 & 0.100 & 0.078 \\
  & 10  & 0.008 & 0.124 & 0.114 & 0.176 & 0.107 \\
  & 50  & 0.010 & 0.145 & 0.080 & 0.133 & 0.097 \\
  & 100 & 0.015 & 0.150 & 0.081 & 0.153 & 0.102 \\
\midrule
\multirow{4}{*}{SpatCode}
  & 1   & 0.998 & 0.938 & 0.752 & 0.956 & 0.982 \\
  & 10  & 1.000 & 0.976 & 0.924 & 0.957 & 0.994 \\
  & 50  & 1.000 & 0.981 & 0.965 & 0.927 & 0.996 \\
  & 100 & 1.000 & 0.974 & 0.966 & 0.886 & 0.992 \\
\bottomrule
\end{tabular}%
}
\end{table}

\subsection{Results and Analysis}
\subsubsection{Latency Comparison}
We evaluate insert and query latency to determine whether SpatCode's single-vector design and circular updates translate into measurable throughput gains when data are arriving continuously. Table \ref{tab:latency} reports the average cost of inserting one record and of retrieving the top-100 neighbors with the HNSW search parameter $e_f=100$. The results show that SpatCode delivers the fastest overall query latency with no insert-time penalty compared to the other Milvus variants. ThalDB appears fastest on insert because only scalar tuples are written (the two image paths are stored as strings), but its query latency exceeds one second because the system must load, decode, and compare both images at run time and then process with the relational filter. This is a workflow optimized for SQL aggregation rather than ANN retrieval. By contrast, the three Milvus-based variants pre-encode every modality and therefore present similar insert costs. Filtered Search is slowest because each row carries four additional scalar columns that must be logged and indexed, and its query latency is much larger than the two vector methods due to the need to scan many extra candidates that satisfy the time and location predicates before the true top-$k$ are reached. Hybrid Search improves query speed by using an ANN index per field, but must still execute four traversals and a final re-ranking. SpatCode further eliminates this overhead: A single traversal of one graph suffices, yielding lower query latency relative to Hybrid Search while inserting marginally faster because only one index is modified.

\begin{figure}[t]
\centering
\includegraphics[width=0.99\columnwidth]{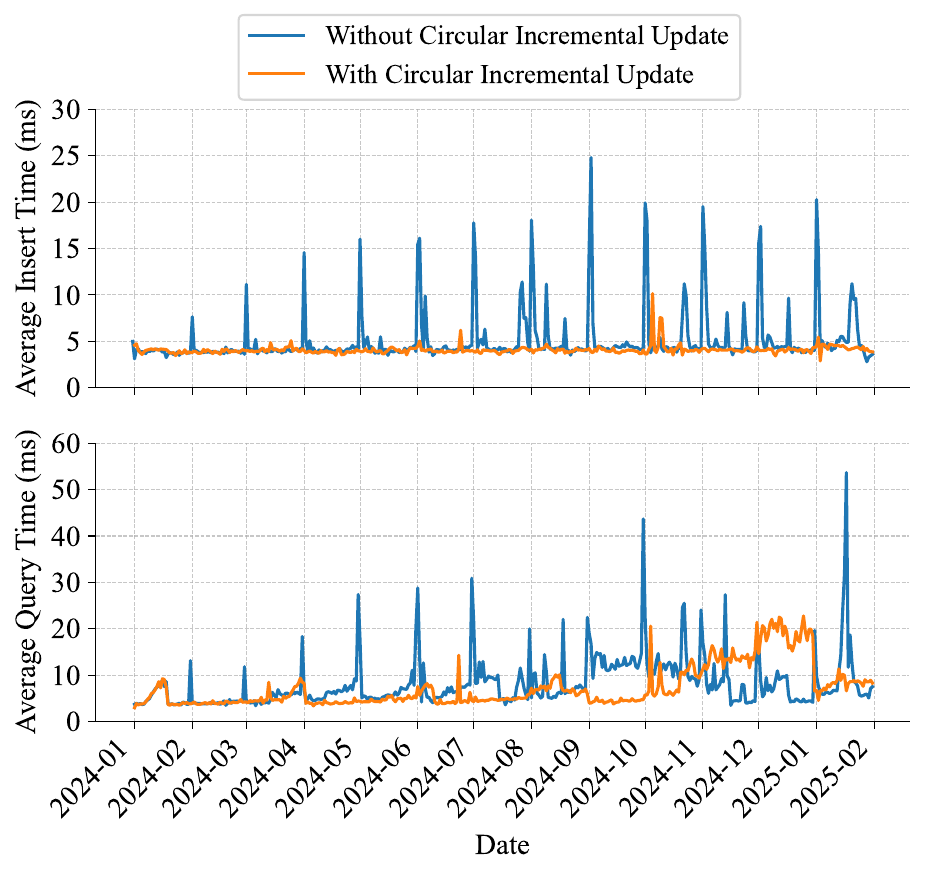}
\caption{Per-record insertion and query latency over thirteen months in the sliding-window.}
\label{fig:time_series}
\end{figure}

\begin{figure}[t]
\centering
\includegraphics[width=0.99\columnwidth]{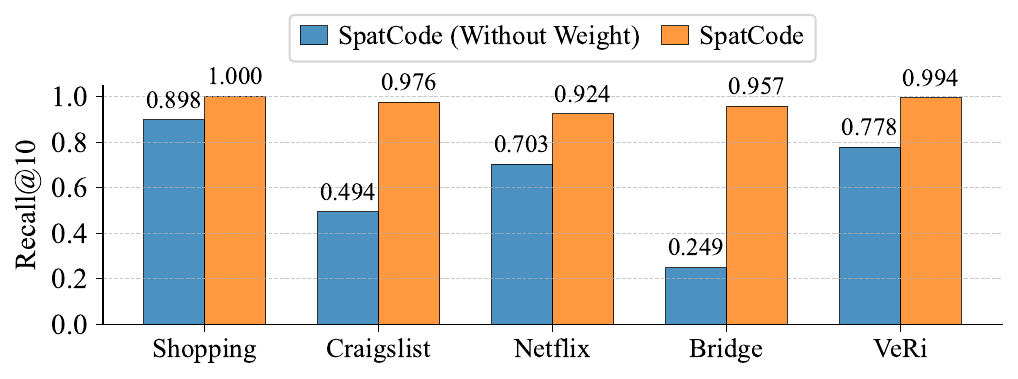}
\caption{Recall@10 of SpatCode with or without weighted interest-based retrieval.}
\label{fig:recall_weight}
\end{figure}

\begin{figure}[t]
\centering
\includegraphics[width=0.99\columnwidth]{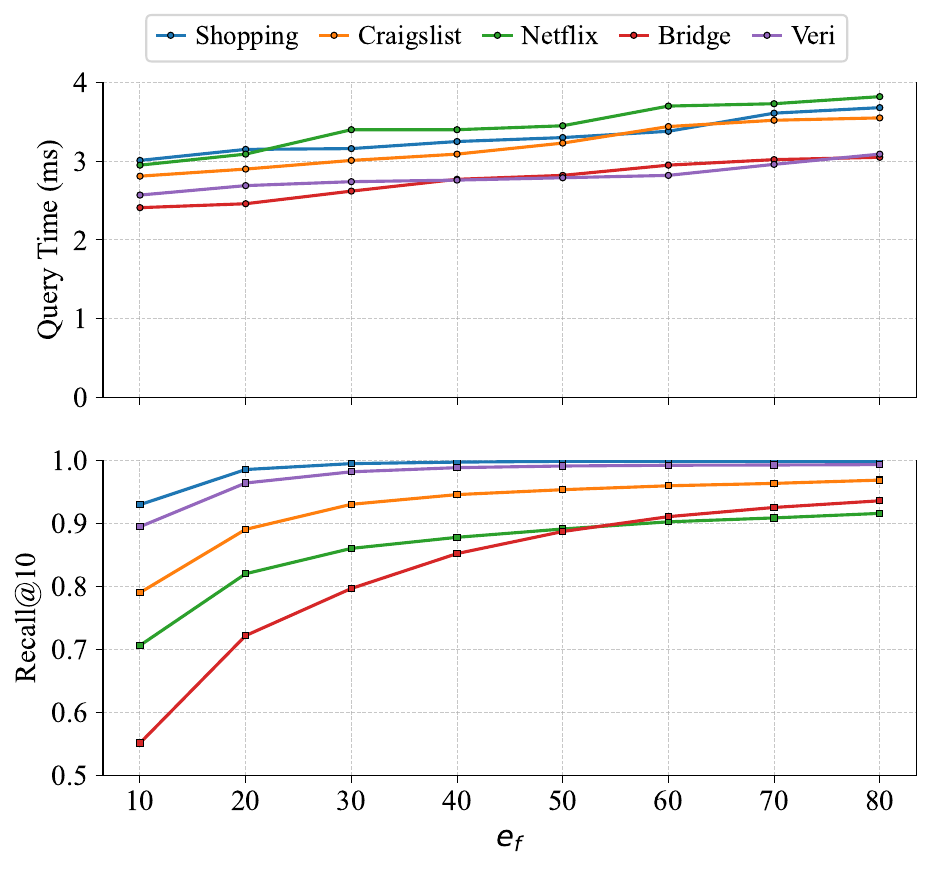}
\caption{Query latency versus recall@10 of SpatCode at various HNSW $e_f$.}
\label{fig:time_recall}
\end{figure}

\begin{figure}[t]
\centering
\includegraphics[width=0.99\columnwidth]{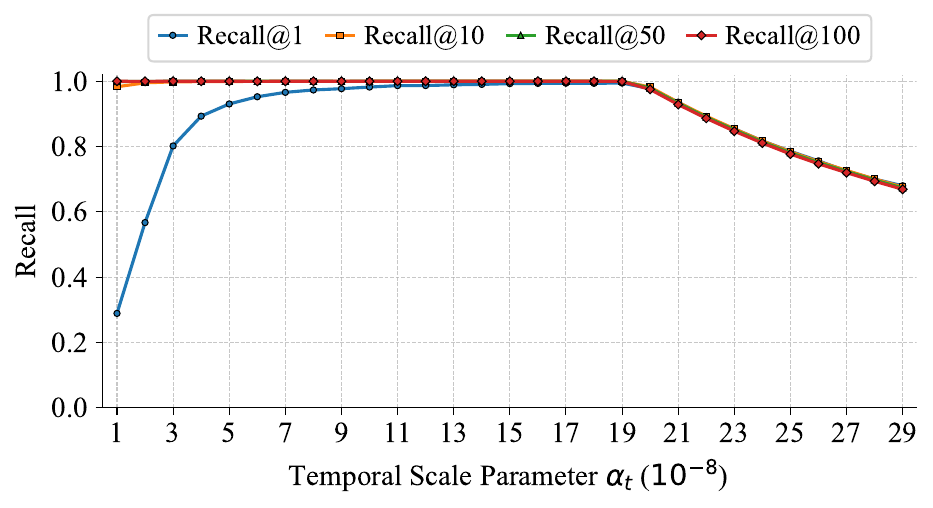}
\caption{Recall versus temporal scale parameter $\alpha_t$ of SpatCode.}
\label{fig:recall_scale_parameter}
\end{figure}

\subsubsection{Retrieval Quality Comparison}
We assess retrieval quality across methods to validate whether SpatCode achieves enhanced performance in fuzzy spatiotemporal retrieval scenarios.
The results show that SpatCode consistently outperforms other methods in recall across every tested cut-off. As summarized in Table \ref{tab:retrieval}, the two hard-filtering baselines (ThalDB and Filtered Search) deliver relatively low effectiveness. Both treat time and location as Boolean constraints rather than graded similarities and use other features alone for ranking, so any record just outside the spatial or temporal window is irretrievably discarded even if its visual content is nearly identical to the query. The recall@$k$ of Hybrid Search never exceeds 0.2 for every $k$, which is the lowest among all methods, demonstrating that many globally relevant items are rejected because they are not in the top local lists of all modalities. SpatCode attains around 0.9 recall@$k$ across all cut-offs, indicating that the weighted distance ranks the correct objects.

\subsubsection{Ablation Study}
To isolate the contribution of the Circular Incremental Update Mechanism, we record insert and query latency over thirteen months with a six-month sliding window. We set $\tau$ to one month and $L=6$. Figures \ref{fig:time_series} contrasts the raw time series with and without the mechanism. The results show that enabling Circular Incremental Update smooths both insert and query latency by eliminating the monthly spikes seen in the naive scheme. Without circular updates, the system must re-encode every timestamp to $[0,\pi)$ at the turn of each month; both insert and query curves exhibit sharp spikes whose amplitude grows with database size, reaching peaks of roughly 25 ms for insert and 50 ms for query by late 2024. After enabling circular updates, the spikes disappear, leaving a near-flat baseline of about 4 ms for insert and query. The residual bumps correspond to background compaction triggered when lazy deletions exceed the fragmentation threshold, yet even these remain below the worst-case pauses observed in the naive scheme. The ablation therefore validates the analytic claim that cyclic bucket retirement yields less incremental maintenance cost and shields online traffic from window-shift churn.

We compare SpatCode with and without interest weights to validate the effectiveness of the Weighted Interest-based Retrieval Algorithm. As shown in Figure \ref{fig:recall_weight}, the weighted variant consistently achieves higher recall@10 across benchmarks. This demonstrates that per-modality interest weights guide the unified similarity measure toward user intent, amplifying informative cues while suppressing distractors. Notably, our approach improves ranking quality without any modification to the underlying encoder. Moreover, by applying weights at query time rather than on the stored vectors, it preserves a single unified index while enabling intent-aware reweighting on the fly, highlighting the method's flexibility and practical efficiency.

Figure \ref{fig:time_recall} presents how the HNSW search parameter $e_f$ affects search accuracy. For SpatCode, sweeping $e_f$ from 10 to 80, recall@10 improves from roughly 0.75 at $e_f=10$ to 0.95 at $e_f=80$, whereas average query time moves from about 2.75 ms to 3.25 ms. The largest gains concentrate in the low-$e_f$ regime and flatten beyond $e_f\ge60$. A practical operating point is $e_f\in[50,60]$, which delivers high recall@10 at near-minimum latency. Across datasets, recall@10 increases smoothly as $e_f$ grows while latency rises modestly, exhibiting the expected trade-off between recall and query time.

We also evaluate how the temporal scale parameter $\alpha_t$ in the rotary time encoding affects retrieval recall. To isolate this effect, we construct a dataset which contains only timestamps sampled at random over a wide horizon. As shown in Figure \ref{fig:recall_scale_parameter}, across the sweep, recall displays a clear unimodal dependence on the temporal scale parameter used in the rotary time encoding. Increasing the scale from very small values initially improves recall by expanding phase differences and enhancing discrimination, but beyond a moderate range recall declines as phase gaps exceed a semicircle. At small scales, temporal phases collapse toward one another under single precision, pushing inner products toward near-indistinguishable values and degrading rank quality. At overly large scales, the encoded phase wraps past the half-circle, so cosine similarity is no longer monotonic in temporal distance, introducing aliasing and harming recall. The evidence supports choosing a scale that lifts typical phase differences above single precision resolution while keeping expected phase gaps strictly below $\pi$, producing recall that best aligns with true temporal proximity under the encoding.

The experiments corroborate the three design hypotheses, confirming that unifying all modalities into a single embedding enables efficient ANN traversal for complex spatiotemporal queries, the weighted interest-based retrieval allows intent-aware reweighting, and the circular incremental update ensures stable long-term performance. Ablations show that improvements persist across datasets and parameter regimes. Together, these results demonstrate that SpatCode provides a practical and efficient solution for real-time spatiotemporal retrieval under rolling time windows and diverse application settings.
% Together, the experiments corroborate the three design hypotheses. First, unifying all modalities into a single embedding allows one ANN traversal to serve complex spatiotemporal queries, cutting end-to-end latency relative to filtered or multi-index baselines. Second, weighted concatenation recovers almost the entire similarity mass, whereas field-wise hybrid retrieval discards candidates that are moderately but jointly relevant. Third, the circular incremental update strategy prevents periodic re-encoding storms, maintaining stable insert and query performance over long-running streams. These results position SpatCode as a practical solution for real-time spatiotemporal retrieval under rolling time windows.

\section{Conclusion}
In this work, we present SpatCode, a unified spatiotemporal vector retrieval framework that addresses challenges of spatiotemporal vector retrieval. SpatCode tightly integrates semantic, temporal, and geospatial cues into a single similarity space via a Rotary-based Unified Encoding Method that preserves cosine-distance monotonicity.
It supports incremental encoding through a Circular Incremental Mechanism and enables personalized, low-latency search with a Weighted Interest-based Retrieval Algorithm. Extensive experiments demonstrate that SpatCode outperforms representative baselines by achieving lower query latency and higher recall. This work lays the foundation for more robust and scalable spatiotemporal vector retrieval systems, addressing the needs of emerging applications.

\begin{acks}
To Robert, for the bagels and explaining CMYK and color spaces.
\end{acks}

\bibliographystyle{ACM-Reference-Format}
\bibliography{sample-base}

\appendix
\section*{Appendix}
\section{Precision Analysis}
We analyze the precision of the time and location encoding, by defining an angular discrimination $\varepsilon_{\cos}$, the minimum cosine decrement needed to declare two embeddings distinct. Typical retrieval engines resolve distances down to $\varepsilon_{\cos}\approx10^{-6}$ without numerical instability in single precision.

Let two events be separated by $\Delta t$. The inner product of their embedding is equal to $\cos(\alpha_t\Delta t)$. To guarantee that any interval exceeding a user-specified threshold $\Delta t_{\min}$ is distinguishable, we impose
$$
|\cos(\alpha_t\Delta t)-1|\ge\varepsilon_{\cos},\ \forall\Delta t\ge \Delta t_{\min}.
$$
Using the small-angle expansion $\cos x\approx1-x^2/2$, the tightest constraint arises at $\Delta t=\Delta t_{\min}$ and yields
$$
\alpha_t\ge\sqrt{\frac{2\varepsilon_{\cos}}{\Delta t_{\min}^2}}.
$$
With $\varepsilon_{\cos}=10^{-6}$ and $\Delta t_{\min}=4$ h ($14400$ s), we obtain $\alpha_t\ge9.82\times10^{-8}$ s$^{-1}$. Equivalently, one full semicircle $\pi$ radians covers
$$
\Delta t_h=\frac{\pi}{\alpha_t}\le3.20\times10^7\ \text{s}=370\ \text{d}.
$$

For two geographic points with central angle $\delta$, the encoder produces inner product $\cos(\delta)$. To resolve any pair separated by at least $\Delta d_{\min}$ kilometers, we translate $\Delta d_{\min}$ to radians via Earth radius $R_{\oplus}=6371$ km:
$$
\delta_{\min}=\frac{\Delta d_{\min}}{R_{\oplus}}.
$$
Reusing the same criterion as the temporal case,
$$
|\cos(\delta)-1|\ge\varepsilon_{\cos},\ \forall\delta\ge\delta_{\min}.
$$
Applying $\cos x\approx1-x^2/2$ yields
$$
\delta_{\min}=\sqrt{2\varepsilon_{\cos}}=1.41\times10^{-3}.
$$
The bound gives $\Delta d_{\min}=9.01$ km, which is the smallest distinguishable distance between two geographic points.

Overall, for the time modality, with the smallest resolvable time difference of 4 hours, the maximum time interval can be 370 days; for the location modality, the smallest distinguishable distance is 9.01 kilometers. When considering double precision, the angular discrimination can be $\varepsilon_{\cos}\approx10^{-15}$, and therefore setting $\Delta t_{\min}=1$ s results in $\Delta t_h=813$ d, while $\Delta d_{\min}=0.285$ m.

\section{Complexity Analysis}
To assess the scalability of SpatCode in large-scale spatiotemporal retrieval, we analyze the computational complexity of query processing.
Let $N$ denote the number of live records, $k$ the number of retrieved results, $d_i$ the dimensionality of the $i$-th modality embedding, and $m$ the number of active modalities. The total embedding dimension for SpatCode is thus $d=\sum_{i=1}^m d_i$. We assume HNSW is used as the underlying approximate nearest neighbor (ANN) method. Query complexity is measured in terms of graph traversals and cosine similarity computations.

We compare three retrieval approaches in terms of query time complexity $T_q$:

\subsection{Scalar-Filtered Vector Search}
This baseline stores a single HNSW graph over multimodal embeddings (excluding time and location), with timestamps $t$ and spatial coordinates $(\lambda, \phi)$ stored as scalar payloads. A query first performs an ANN search, visiting enough nodes to find $k$ valid candidates and computing an equal number of cosine similarities.

Assuming a stationary data stream and a temporal query window $\Delta t_h$, the expected probability that a candidate survives this filtering step is $p_h = \Delta t_h / \Delta t^*$, where $\Delta t^*$ is the global time span of the dataset. Since $\Delta t_h$ often spans only a small proportion of $\Delta t^*$, in most cases $p_h\ll1$. If the vectors are uniformly distributed, the expected number of candidates will be $k/p_h$. When $p_h \ll 1$, the number of candidates must increase proportionally to ensure sufficient recall, increasing the cost accordingly. The overall query complexity is:
$$T_q=O(\log N+k/p_h).$$

\subsection{Hybrid Multi-Field Vector Search}
This baseline constructs separate HNSW graphs for each modality, including time and location, treated as independent vectors. Each modality-specific search visits $O(\log N)$ nodes and retrieves a candidate buffer of size $k_i$ (assumed equal to $k$ across modalities for simplicity). A merging step then computes a weighted sum of similarities
$$S=\sum_{i=1}^mw_i\langle v_i,q_i\rangle$$
across all distinct candidates. The total query complexity includes independent traversals and merging over $m$ modalities:
$$T_q=O(m(\log N+k)).$$

\subsection{SpatCode (Ours)}
SpatCode encodes all modalities—including time and location—into a unified $d$-dimensional unit-normalized vector $\tilde{v}$. All records are stored in a single HNSW index using this concatenated representation. A query forms $\tilde{q} = [w_1 q_1;\dots; w_m q_m]$, ensuring consistency with the database encoding. Since modality weights are embedded at the vector construction stage, no post-filtering or merging is required. A single cosine similarity is computed per visited node. The query complexity is:
$$T_q=O(\log N+k).$$

In typical settings where $p_h\ll1$ and $m>1$, we have:
$$
T_q^{\text{SpatCode}}<T_q^{\text{Hybrid}}<T_q^{\text{Filtered}}.
$$
Among the three approaches, SpatCode achieves the lowest query complexity. For Filtered Search, the expected complexity is $T_q=O(\log N+k/p_h)$, where the filter survival probability $p_h\ll1$ under tight temporal or spatial constraints, causing $k/p_h\gg k$, and thus making it the slowest method. Hybrid Search avoids scalar filtering but requires $m$ independent traversals and merging, resulting in complexity $T_q=O(m(\log N+k))$, which grows linearly with the number of modalities $m$. In contrast, SpatCode requires only a single HNSW traversal and similarity computation, achieving complexity $T_q=O(\log N+k)$.

\end{document}